\documentclass[12pt] {article} 

\usepackage{psfrag}
\usepackage{graphicx}
\usepackage{dcolumn}
\usepackage{latexsym,amsfonts}
\usepackage{bm}
\usepackage{amssymb}
\begin{document}
\setcounter{page}{1}
\def\theequation{\arabic{section}.\arabic{equation}}
\def\theequation{\thesection.\arabic{equation}}
\setcounter{section}{0}

\title{On renormalizability of the massless Thirring
  model}

\author{H. Bozkaya\,\thanks{E--mail: hidir@kph.tuwien.ac.at, Tel.:
+43--1--58801--14262, Fax: +43--1--58801--14299}\,,
A. N. Ivanov\,\thanks{E--mail: ivanov@kph.tuwien.ac.at, Tel.:
+43--1--58801--14261, Fax: +43--1--58801--14299}~\thanks{Permanent
Address: State Polytechnic University, Department of Nuclear
Physics, 195251 St. Petersburg, Russian Federation}\,, 
M. Pitschmann\,\thanks{E--mail: pitschmann@kph.tuwien.ac.at, Tel.:
+43--1--58801--14263, Fax: +43--1--58801--14299}}

\date{\today}

\maketitle

\begin{center} {\it Atominstitut der \"Osterreichischen
    Universit\"aten, Technische Universit\"at Wien,\\ Wiedner Hauptstrasse
    8-10, A-1040 Wien, \"Osterreich }
\end{center}

\begin{center}
\begin{abstract}
  We discuss the renormalizability of the massless Thirring model in
  terms of the causal fermion Green functions and correlation
  functions of left--right fermion densities. We obtain the most
  general expressions for the causal two--point Green function and
  correlation function of left--right fermion densities with dynamical
  dimensions of fermion fields, parameterised by two parameters. The
  region of variation of these parameters is constrained by the
  positive definiteness of the norms of the wave functions of the
  states related to components of the fermion vector current.  We show
  that the dynamical dimensions of fermion fields calculated for
  causal Green functions and correlation functions of left--right
  fermion densities can be made equal. This implies the
  renormalizability of the massless Thirring model in the sense that
  the ultra--violet cut--off dependence, appearing in the causal
  fermion Green functions and correlation functions of left--right
  fermion densities, can be removed by renormalization of the wave
  function of the massless Thirring fermion fields only.  
\end{abstract}
\end{center}

\newpage

\section{Introduction}
\setcounter{equation}{0}

The massless Thirring model \cite{WT58} is an exactly solvable quantum
field theoretic model of fermions with a non--trivial four-fermion
interaction in 1+1--dimensional space--time defined by the Lagrangian
invariant under the chiral group $U_{\rm V}(1)\times U_{\rm A}(1)$
\begin{eqnarray}\label{label1.1}
{\cal L}_{\rm Th}(x) = \bar{\psi}(x)i\gamma^{\mu}\partial_{\mu}\psi(x) -
\frac{1}{2}\,g\,\bar{\psi}(x)\gamma^{\mu}\psi(x)\bar{\psi}(x)
\gamma_{\mu}\psi(x),
\end{eqnarray}
where $\psi(x)$ is a massless Dirac fermion field and $g$ is a
dimensionless coupling constant that can be both positive and
negative.

A solution of the Thirring model assumes a development of a procedure
for the calculation of any correlation function
\cite{KJ61}--\cite{RO81}. As has been shown by Hagen \cite{CH67} and
Klaiber \cite{BK67}, the correlation functions of massless Thirring
fermion fields can be parameterised by one arbitrary parameter. In
Hagen's notation this parameter is $\xi$.  Below we show that the
correlation functions in the massless Thirring model can be
parameterised by two parameters (see Appendix A). This
confirms the results obtained by Harada {\it et al.}  \cite{KH86} (see
also \cite{RJ84,GC83}) for the chiral Schwinger model. In our notation
these parameters are $\bar{\xi}$ and $\bar{\eta}$. The region of
variation of these parameters is restricted by the condition for the
norms of the wave functions of the states related to the components of
the fermion vector current to be positive (see Appendix B).  For
$\bar{\eta} = 1$ the parameter $\bar{\xi}$ is equal to Hagen's
parameter $\bar{\xi} = \xi$.  The parameters $\bar{\xi}$ and
$\bar{\eta}$ we use for the analysis of the non--perturbative
renormalizability of the massless Thirring model in the sense that a
dependence of any correlation function on the ultra--violet cut--off
$\Lambda$ can be removed by the renormalization of the wave function
of Thirring fermion fields only.  We show that independence of any
correlation function of an ultra--violet cut--off exists only if the
dynamical dimensions of Thirring fermion fields, calculated from
different correlation functions, are equal. We would like to remind
that for the known solutions of the massless Thirring model
\cite{KJ61}--\cite{RO81} the dynamical dimensions of massless Thirring
fermion fields, calculated from causal Green functions and left--right
correlation functions, are different.  The existence of different
dynamical dimensions of Thirring fermion fields obtained from
different correlation functions has been regarded by Jackiw as a
problem of 1+1--dimensional quantum field theories \cite{RJ71}.

The paper is organised as follows. In Section 2 we define the
generating functional of correlation functions in the massless
Thirring model. In Section 3 we calculate the two--point causal Green
function and the dynamical dimension of massless Thirring fermion
fields $d_{\bar{\psi}\psi}(g)$, parameterised by two parameters
$\bar{\xi}$ and $\bar{\eta}$. In Section 4 we calculate the two--point
correlation function of the left--right fermion densities and the
dynamical dimension of massless Thirring fermion fields
$d_{(\bar{\psi}\psi)^2}(g)$ in dependence on $\bar{\xi}$ and
$\bar{\eta}$.  We show that the dynamical dimensions
$d_{\bar{\psi}\psi}(g)$ and $d_{(\bar{\psi}\psi)^2}(g)$ can be made
equal $d_{(\bar{\psi}\psi)^2}(g) = d_{\bar{\psi}\psi}(g)$.  This
indicates that the massless Thirring model is renormalizable in the
sense that the dependence of the causal Green functions and
correlation functions of left--right fermion densities on the
ultra--violet cut--off can be removed by the renormalization of the
wave function of massless Thirring fermion fields.  In Section 5 we
corroborate the validity of this assertion within the standard
renormalization procedure \cite{JC84}.  In the Conclusion we discuss
the obtained results. In Appendix A we show that the determinant ${\rm
  Det}(i\hat{\partial} + \hat{A})$, where $A_{\mu}$ is an external
vector field, can be parameterised by two parameters. For
this aim we calculate the vacuum expectation value of the vector
current and show that the ambiguous parameterisation of the
determinant ${\rm Det}(i\hat{\partial} + \hat{A})$ is fully caused by
the regularization procedure \cite{KH86}--\cite{GC83}. In Appendix B
we analyse the constraints on the parameters $\bar{\xi}$ and
$\bar{\eta}$ imposed by the positive definiteness of the norms of the
wave functions of the states related to the components of the vector
fermion current. We show that the positive definiteness of these norms
does not prohibit the possibility for the dynamical dimensions of
massless Thirring fermion fields to be equal.  According to the
equivalence of the massive Thirring model to the sine--Gordon model
\cite{SC75}, the constraints on the parameters $\bar{\eta}$ and
$\bar{\xi}$ together with the requirement of the non--perturbative
renormalizability of the massless Thirring model lead to the strongly
coupled sine--Gordon field with the coupling constant $\beta^2 \sim
8\pi$. A behaviour and renormalizability of the sine--Gordon model for
the coupling constants $\beta^2 \sim 8\pi$ has been investigated in
\cite{SGM}.

\section{Generating functional of correlation functions}
\setcounter{equation}{0}

The generating functional of vacuum expectation values of products of
massless Thirring fermion fields, i.e. correlation functions, is
defined by
\begin{eqnarray}\label{label2.1}
  Z_{\rm Th}[J,\bar{J}] &=&\int {\cal D}\psi{\cal
    D}\bar{\psi}\,\exp\,i\int
  d^2x\,\Big[\bar{\psi}(x)i\gamma^{\mu}\partial_{\mu}\psi(x) -
  \frac{1}{2}\,g\,\bar{\psi}(x)\gamma^{\mu}\psi(x)\bar{\psi}(x)\nonumber\\
  &&+ \bar{\psi}(x)J(x) +\bar{J}(x)\psi(x)\Big].
\end{eqnarray}
It can be represented also as follows 
\begin{eqnarray}\label{label2.2}
Z_{\rm Th}[J,\bar{J}] = \exp\Big\{\,\frac{i}{2}\,g\int
d^2x\,\frac{\delta}{\delta A_{\mu}(x)}\frac{\delta}{\delta
A^{\mu}(x)}\Big\}Z^{(0)}_{\rm Th}[A; J,\bar{J}]\Big|_{A = 0},
\end{eqnarray}
where we have denoted
\begin{eqnarray}\label{label2.3}
  Z^{(0)}_{\rm Th}[A; J,\bar{J}] &=& \int {\cal D}\psi{\cal
  D}\bar{\psi}\,\exp\,i\int
  d^2x\,\Big[\bar{\psi}(x)i\gamma^{\mu}\partial_{\mu}\psi(x) +
  \bar{\psi}(x)\gamma^{\mu}\psi(x)A_{\mu}(x)\nonumber\\ &&+
  \bar{\psi}(x)J(x) + \bar{J}(x)\psi(x)\Big].
\end{eqnarray}
The functional $Z^{(0)}_{\rm Th}[A; J,\bar{J}]$ is a generating
functional of vacuum expectation values of products of massless
fermion fields of the massless Schwinger model coupled to an external
vector field $A_{\mu}(x)$ \cite{JS62}. The integration over fermion
fields can be carried out explicitly and we get
\begin{eqnarray}\label{label2.4}
  Z^{(0)}_{\rm Th}[A; J,\bar{J}] = {\rm Det}(i\hat{\partial} +
  \hat{A})\,\exp\Big\{i\,\int\!\!\!\int
  d^2x\,d^2y\,\bar{J}(x)\,G(x,y)_A\,J(y)\Big\},
\end{eqnarray}
where $G(x,y)_A$ is a two--point causal fermion Green function obeying
the equation
\begin{eqnarray}\label{label2.5}
  i\gamma^{\mu}\Big(\frac{\partial}{\partial x^{\mu}} -
  i\,A_{\mu}(x)\Big)G(x,y)_A = -\,\delta^{(2)}(x - y).
\end{eqnarray}
As has been shown in the Appendix A, the functional determinant ${\rm
  Det}(i\hat{\partial} + \hat{A})$ can be parameterised by two
parameters
\begin{eqnarray}\label{label2.6}
{\rm Det}(i\hat{\partial} + \hat{A}) = \exp\Big\{\frac{i}{2}\int\!\!\!\int d^2x
d^2y\,A_{\mu}(x)\,D^{\mu\nu}(x - y)\,A_{\nu}(y)\Big\},
\end{eqnarray}
where we have denoted 
\begin{eqnarray}\label{label2.7}
  D^{\mu\nu}(x - y) = \frac{\bar{\xi}}{\pi}\,g^{\mu\nu}\,\delta^{(2)}(x - y) -
  \frac{\bar{\eta}}{\pi}\,\,\frac{\partial}{\partial
    x_{\mu}}\frac{\partial}{\partial x_{\nu}}\Delta(x - y; \mu).
\end{eqnarray}
Here $\bar{\xi}$ and $\bar{\eta}$ are two parameters,
$g^{\mu\nu}$ is the metric tensor and $\Delta(x - y; \mu)$ is the
causal two--point Green function of a free massless (pseudo)scalar
field
\begin{eqnarray}\label{label2.8}
i\Delta(x - y; \mu) = \frac{1}{4\pi}\,
{\ell n}[ - \mu^2(x - y)^2 + i\,0].
\end{eqnarray}
It obeys the equation $\Box_x\Delta(x - y; \mu) = \delta^{(2)}(x -
y)$, where $\mu$ is an infrared cut--off.

The appearance of two parameters is caused by dependence of the
calculation of the determinant ${\rm Det}(i\hat{\partial} + \hat{A})$
on the regularization procedure \cite{KH86}--\cite{GC83}. In Appendix
B we find the constraint on the region of variation of these
parameters imposed by the positive definiteness of the norms of the
wave functions of the states related to the components of the fermion
vector current. The parameters $\bar{\xi}$ and $\bar{\eta}$ are
related to Hagen's parameter $\xi$ as $\bar{\xi} = \xi$ and
$\bar{\eta} = 1$. 

The solution of the equation (\ref{label2.5}) is
equal to
\begin{eqnarray}\label{label2.9}
\hspace{-0.3in}&&G(x,y)_A = G_0(x - y)\nonumber\\
\hspace{-0.3in}&&\times\,\exp\Big\{-i\,(g^{\alpha\beta} -
\varepsilon^{\alpha\beta}\,\gamma^5)\int
d^2z\,\frac{\partial}{\partial z^{\alpha}}[\Delta(x - z; \mu) -
\Delta(y - z; \mu)]\,A_{\beta}(z)\Big\},
\end{eqnarray}
where $\varepsilon^{\alpha\beta}$ is the antisymmetric tensor defined
by $\varepsilon^{01} = 1$ and $G_0(x - y)$ is the Green function of a
free massless fermion field
\begin{eqnarray}\label{label2.10}
G_0(x - y) = i\gamma^{\mu}\frac{\partial}{\partial x^{\mu}}\Delta(x -
y; \mu) = \frac{1}{2\pi}\,\frac{\gamma^{\mu}(x - y)_{\mu}}{(x - y)^2 -
i\,0}
\end{eqnarray}
satisfying the equation $i\gamma^{\mu}\partial_{\mu}G_0(x - y) = -
\,\delta^{(2)}(x - y)$.

Any correlation function of the massless Thirring fermion fields can
be defined by functional derivatives of the generating functional
(\ref{label2.1}) and calculated in terms of the two--point Green
functions $G(x,y)_A$ and $\Delta(x - y; \mu)$.  Below we calculate the
casual two--point Green function $G(x,y)$ and the correlation function
$C(x,y)$ of the left--right fermion densities defined by
\begin{eqnarray}\label{label2.11}
  \hspace{-0.3in} G(x,y) &=& i\,\langle 0|{\rm
    T}(\psi(x)\bar{\psi}(y))|0\rangle = \frac{1}{i}\,\frac{\delta}{\delta
    \bar{J}(x)}\,\frac{\delta}{\delta J(y)}Z_{\rm Th}[J, \bar{J}]\Big|_{J= \bar{J} = 0},
  \nonumber\\
  \hspace{-0.3in}  C(x,y) &=&\Big\langle 0\Big|{\rm
    T}\Big(\bar{\psi}(x)\Big(\frac{1-\gamma^5}{2}\Big)
  \psi(x)\,\bar{\psi}(y)\Big(\frac{1+\gamma^5}{2}\Big)
  \psi(y)\Big)\Big|0\Big\rangle = \nonumber\\
  \hspace{-0.3in} &=& \frac{1}{i}\frac{\delta}{\delta J(x)}
  \Big(\frac{1 - \gamma^5}{2}\Big)
  \frac{1}{i}\frac{\delta}{\delta \bar{J}(x)}
  \frac{1}{i}\frac{\delta}{\delta J(y)}\Big(\frac{1 + \gamma^5}{2}\Big)
  \frac{1}{i}\frac{\delta}{\delta \bar{J}(y)}Z_{\rm Th}[J,\bar{J}]\Big|_{J = \bar{J} = 0},
\end{eqnarray}
where ${\rm T}$ is the time--ordering operator.  The main aim of the
investigation of these correlation functions is in the calculation of
the dynamical dimensions of the massless Thirring fermion fields and
the analysis of the possibility to make them equal \cite{RJ71}.

\section{Two--point causal Green function $G(x,y)$}
\setcounter{equation}{0}

In terms of the generating functional (\ref{label2.1}) the two--point
Green function $G(x,y)$ is defined by
\begin{eqnarray}\label{label3.1}
&&G(x,y) = \frac{1}{i}\,\frac{\delta}{\delta
\bar{J}(x)}\,\frac{\delta}{\delta J(y)}Z_{\rm Th}[J, \bar{J}]\Big|_{J= \bar{J} = 0} = 
\exp\Big\{\,\frac{i}{2}\,g\int
d^2z\,\frac{\delta}{\delta A_{\mu}(z)}\frac{\delta}{\delta
A^{\mu}(z)}\Big\}\nonumber\\ &&\hspace{0.5cm}\times\,\exp\Big\{\frac{i}{2}\int\!\!\!\int
d^2z_1 d^2z_2\,A_{\lambda}(z_1)\,D^{\lambda\varphi}(z_1 -
z_2)\,A_{\varphi}(z_2)\Big\}\,G(x,y)_A\Big|_{A = 0}.
\end{eqnarray}
The calculation of the r.h.s. of (\ref{label3.1}) reduces to the
calculation of the path integral
\begin{eqnarray}\label{label3.2}
  G(x,y)&=& \frac{1}{2\pi}\,\frac{\gamma^{\mu}(x - y)_{\mu}}{(x -
    y)^2 - i\,0}\int {\cal D}^2u\,\exp\Big\{-\,\frac{i}{2}\int
  d^2z\,u_{\mu}(z)u^{\mu}(z)\nonumber\\ && -
  \frac{i}{2}\,g\int\!\!\!\int
  d^2z_1d^2z_2\,u_{\mu}(z_1)\,D^{\mu\nu}(z_1 - z_2)\,u_{\nu}(z_2) +
  \,\sqrt{g}\,(g^{\alpha\beta} -
  \varepsilon^{\alpha\beta}\,\gamma^5)\nonumber\\ &&\times \int
  d^2z\,\frac{\partial}{\partial z^{\alpha}}[\Delta(x - z; \mu) -
  \Delta(y - z; \mu)]\,u_{\beta}(z)\Big\}.
\end{eqnarray}
Symbolically the r.h.s. of (\ref{label3.2}) can be written as
\begin{eqnarray}\label{label3.3}
\hspace{-0.3in}&&G(x,y) = \frac{1}{2\pi}\,\frac{\gamma^{\mu}(x -
y)_{\mu}}{(x - y)^2 - i\,0}\nonumber\\ 
\hspace{-0.3in}&&\times\int
{\cal D}^2u\,\exp\Big\{ - \frac{i}{2}\,u(1 + gD)u +
\sqrt{g}\,\partial(\Delta_x -\Delta_y)\,u -
\sqrt{g}\,\gamma^5\,\partial(\Delta_x
-\Delta_y)\,\varepsilon\,u\Big\}.
\end{eqnarray}
The integration over $u$ can be carried out by quadratic extension.
This yields
\begin{eqnarray}\label{label3.4}
G(x,y) &=& \frac{1}{2\pi}\,\frac{\gamma^{\mu}(x - y)_{\mu}}{(x -
y)^2 - i\,0}\,\exp\Big\{ -\,\frac{i}{2}\,g\,\partial(\Delta_x -
\Delta_y)\,\frac{1}{1 + gD}\,\partial(\Delta_x - \Delta_y)\nonumber\\
&& + \frac{i}{2}\,g\,\partial(\Delta_x -
\Delta_y)\,\varepsilon\,\frac{1}{1 +
gD}\,\varepsilon\,\partial(\Delta_x - \Delta_y)\Big\}.
\end{eqnarray}
For the subsequent calculation we have to construct the matrix $(1 +
gD)^{-1}$. The matrix $(1 + gD)$ has the following elements
\begin{eqnarray}\label{label3.5}
  (1 + gD)^{\mu\alpha}(x,z) =  \Big(1 +
  \bar{\xi}\,\frac{g}{\pi}\Big)\,g^{\mu\alpha}\,\delta^{(2)}(x - z) -
  \bar{\eta}\,\frac{g}{\pi}\,\frac{\partial}{\partial
    x_{\mu}}\frac{\partial}{\partial x_{\alpha}}\Delta(x - z;\mu).
\end{eqnarray}
The elements of the matrix $(1 + gD)^{-1}$ we define as
\begin{eqnarray}\label{label3.6}
  ((1 + gD)^{-1})_{\alpha\nu}(z,y) =  A\,g_{\alpha\nu}\,\delta^{(2)}(z - y) + 
  B\,
  \frac{\partial}{\partial z^{\alpha}}\frac{\partial}{\partial
    z^{\nu}}\Delta(z - y;\mu).
\end{eqnarray}
The matrices $(1 + gD)$ and $(1 + gD)^{-1}$ should obey the condition
\begin{eqnarray}\label{label3.7}
\int d^2z\,(1 + gD)^{\mu\alpha}(x,z)((1 + gD)^{-1})_{\alpha\nu}(z,y) =
g^{\mu}_{\nu}\,\delta^{(2)}(x - y).
\end{eqnarray}
This gives
\begin{eqnarray}\label{label3.8}
  ((1 + gD)^{-1})_{\alpha\nu}(z,y)&=& 
  \frac{g_{\alpha\nu}}{\displaystyle 1 +
    \bar{\xi}\,\frac{g}{\pi}}\,\delta^{(2)}(z - y)\nonumber\\
  &&+ 
  \frac{g}{\pi}\,\frac{\bar{\eta}}{\displaystyle \Big(1 +
    \bar{\xi}\,\frac{g}{\pi}\Big)\Big(1 +
    (\bar{\xi} - \bar{\eta})\,\frac{g}{\pi}\Big)}\,\frac{\partial}{\partial
    z^{\alpha}}\frac{\partial}{\partial z^{\nu}}\Delta(z - y;\mu).
\end{eqnarray}
Using (\ref{label3.8}) we obtain
\begin{eqnarray}\label{label3.9}
  \hspace{-0.3in}-\,\frac{i}{2}\,g\,\partial(\Delta_x - \Delta_y)\,\frac{1}{1 +
    gD}\,\partial(\Delta_x - \Delta_y) &=&   \frac{g}{\displaystyle 1+
    (\bar{\xi} - \bar{\eta})\,\frac{g}{\pi}}\,[i\Delta(0; \mu) - i\Delta(x - y;
  \mu)],\nonumber\\ 
  \hspace{-0.3in}+ \frac{i}{2}\,g\,\partial(\Delta_x -
  \Delta_y)\,\varepsilon\,\frac{1}{1 +
    gD}\,\varepsilon\,\partial(\Delta_x - \Delta_y) &=& -\,\frac{g}{\displaystyle 1+
    \bar{\xi}\,\frac{g}{\pi}}
  [i\Delta(0; \mu) -
  i\Delta(x - y; \mu)],\nonumber\\ 
  \hspace{-0.3in}
\end{eqnarray}
where $i\Delta(0; \mu)$ is equal to
\begin{eqnarray}\label{label3.10}
i\Delta(0; \mu) = -\,\frac{1}{4\pi}\,{\ell
n}\Big(\frac{\Lambda^2}{\mu^2}\Big).
\end{eqnarray}
Thus, the two--point Green function reads
\begin{eqnarray}\label{label3.11}
  G(x,y) &=& \frac{1}{2\pi}\,\frac{\gamma^{\mu}(x - y)_{\mu}}{(x -
    y)^2 - i\,0}\,e^{\textstyle\,4\pi\,d_{\bar{\psi}\psi}(g)
    \,[i\Delta(0; \mu) - i\Delta(x - y; \mu)]} \nonumber\\
  &=&-\, \frac{\Lambda^2}{2\pi}\,\frac{\gamma^{\mu}(x - y)_{\mu}}{
    -\Lambda^2(x - y)^2 + i\,0}\,
  [-\Lambda^2(x - y)^2 + i\,0\,]^{-\,d_{
      (\bar{\psi}\psi)}(g)} \nonumber\\
  &=& \Lambda\,G(d_{\bar{\psi}\psi}(g); \Lambda x,\Lambda y),
\end{eqnarray}
where $d_{\bar{\psi}\psi}(g)$ is a dynamical dimension of the Thirring
fermion field defined by \cite{RJ71}
\begin{eqnarray}\label{label3.12}
  d_{\bar{\psi}\psi}(g) = \frac{g^2}{4\pi^2}\,\frac{\bar{\eta}}{\displaystyle \Big(1 +
    \bar{\xi}\,\frac{g}{\pi}\Big)\Big(1 +
    (\bar{\xi} - \bar{\eta})\,\frac{g}{\pi}\Big)}.
\end{eqnarray}
Now we are proceeding to the calculation of the correlation function
$C(x,y)$.

\section{Two--point correlation function $C(x,y)$}
\setcounter{equation}{0}

According to Eq.(\ref{label2.11}), the two--point correlation function
$C(x,y)$ of the left--right fermion densities is defined by
\begin{eqnarray}\label{label4.1}
  \hspace{-0.3in}&&C(x,y) = \frac{1}{i}\frac{\delta}{\delta J(x)}
\Big(\frac{1 - \gamma^5}{2}\Big)
  \frac{1}{i}\frac{\delta}{\delta \bar{J}(x)}
  \frac{1}{i}\frac{\delta}{\delta J(y)}\Big(\frac{1 + \gamma^5}{2}\Big)
  \frac{1}{i}\frac{\delta}{\delta \bar{J}(y)}Z_{\rm Th}[J,\bar{J}]\Big|_{J = \bar{J} = 0}=
  \nonumber\\
    \hspace{-0.3in}&&=  -\,\exp\Big\{\,\frac{i}{2}\,g\int
  d^2z\,\frac{\delta}{\delta A_{\mu}(z)}\frac{\delta}{\delta
    A^{\mu}(z)}\Big\}\exp\Big\{\frac{i}{2}\int\!\!\!\int
  d^2z_1 d^2z_2\,A_{\lambda}(z_1)\,D^{\lambda\varphi}(z_1 -
  z_2)\,A_{\varphi}(z_2)\Big\}\nonumber\\  \hspace{-0.3in}&&\times\,{\rm
    tr}\Big\{G(y,x)_A\Big(\frac{1 - \gamma^5}{2}\Big)G(x,y)_A\Big(\frac{1
    + \gamma^5}{2}\Big)\Big\}\Big|_{A = 0}.
\end{eqnarray}
This reduces to the calculation of the path integral
\begin{eqnarray}\label{label4.2}
\hspace{-0.3in}C(x,y) &=& \frac{1}{4\pi^2}\,\frac{1}{(x - y)^2 -
i\,0}\nonumber\\ \hspace{-0.3in}&&\times\,\int {\cal D}^2u\,\exp\Big\{
- \frac{i}{2}\,u(1 + gD)u - 2\,\sqrt{g}\,\partial(\Delta_x
-\Delta_y)\,\varepsilon\,u\Big\}=\nonumber\\ \hspace{-0.3in}&=&
\frac{1}{4\pi^2}\,\frac{1}{(x - y)^2 -
i\,0}\,\exp\Big\{2\,i\,g\,\partial(\Delta_x
-\Delta_y)\,\varepsilon\,\frac{1}{1 + g
D}\,\varepsilon\,\partial(\Delta_x -\Delta_y)\Big\}
\end{eqnarray}
The result is
\begin{eqnarray}\label{label4.3}
  C(x,y) &=& \frac{1}{4\pi^2}\,\frac{1}{(x - y)^2 - i\,0}\,
  e^{\textstyle\, 8\,\pi\,d_{(\bar{\psi}\psi)^2}\,[i\Delta(0; \mu) -
    i\Delta(x - y; \mu)]} = \nonumber\\
  &=&-\,\frac{\Lambda^2}{4\pi^2}\,\frac{1}{-\,\Lambda^2 (x - y)^2 + i\,0}\,
[-\,\Lambda^2 (x - y)^2 + i\,0]^{\textstyle -\,2d_{(\bar{\psi}\psi)^2}} =\nonumber\\
&=& \Lambda^2\,C(d_{(\bar{\psi}\psi)^2}(g); \Lambda x, \Lambda y).
\end{eqnarray}
The dynamical dimension $d_{(\bar{\psi}\psi)^2}$ is equal to 
\begin{eqnarray}\label{label4.4}
  d_{(\bar{\psi}\psi)^2}(g) = - \frac{g}{2\pi}\, \frac{1}{\displaystyle 1 +
    \bar{\xi}\,\frac{g}{\pi}}.
\end{eqnarray}
For $\bar{\xi}$ and $\bar{\eta}$, restricted only by the constraint
caused by the positive definiteness of the norms of the wave functions
of the states related to the components of the fermion vector current
(see Appendix B), the dynamical dimensions of the massless Thirring
model, calculated for the two--point causal Green function
(\ref{label3.12}) and the correlation function of the left--right
fermion densities (\ref{label4.4}), are not equal.  According to
Jackiw \cite{RJ71}, this is a problem of quantum field theories in
1+1--dimensional space--time. However, equating
$d_{(\bar{\psi}\psi)^2}(g)$ and $d_{\bar{\psi}\psi}(g)$ we get the
constraint on the parameter $\bar{\eta}$
\begin{eqnarray}\label{label4.5}
  \bar{\eta} = \frac{2\pi}{g}\,\Big(1 +
    \bar{\xi}\,\frac{g}{\pi}\Big).
\end{eqnarray}
As has been shown in Appendix B, the constraint on the region of
variation of parameters $\bar{\xi}$ and $\bar{\eta}$, imposed by the
positive definiteness of the norms of the wave functions of the states
related to the components of the vector current, does not prevent from
the equality of dynamical dimensions $d_{(\bar{\psi}\psi)^2}(g) =
d_{\bar{\psi}\psi}(g)$.

This indicates that the massless Thirring model is renormalizable.
The dependence on the ultra--violet cut--off $\Lambda$ can be removed
by the renormalization of the wave functions of Thirring fermion
fields either for the $2n$--point Green functions $G(x_1,\ldots,x_n;
y_1, \ldots,y_n)$ or for the $2n$--point correlation functions
$C(x_1,\ldots,x_n; y_1, \ldots,y_n)$ of the left--right fermion
densities. The dynamical  dimension of the Thirring fermion fields is
equal to $d_{\psi}(g) = d_{(\bar{\psi}\psi)^2}(g)$ defined by
Eq.(\ref{label4.4}).

\section{Non--perturbative renormalization}
\setcounter{equation}{0}

According to the standard procedure of renormalization in quantum
field theory \cite{JC84} the renormalizability of the massless
Thirring model should be understood as a possibility to remove all
ultra--violet and infrared divergences by renormalization of the wave
function of the massless Thirring fermion field $\psi(x)$ and the
coupling constant $g$.

Let us rewrite the Lagrangian (\ref{label1.1}) in terms of {\it bare}
quantities
\begin{eqnarray}\label{label5.1}
{\cal L}_{\rm Th}(x) =
\bar{\psi}_0(x)i\gamma^{\mu}\partial_{\mu}\psi_0(x) -
\frac{1}{2}\,g_0\,\bar{\psi}_0(x)\gamma^{\mu}\psi_0(x)\bar{\psi}_0(x)
\gamma_{\mu}\psi_0(x),
\end{eqnarray}
where $\psi_0(x)$, $\bar{\psi}_0(x)$ are {\it bare} fermionic field
operators and $g_0$ is a {\it bare} coupling constant.

The renormalized Lagrangian ${\cal L}(x)$ of the massless Thirring
model should then read \cite{JC84}
\begin{eqnarray}\label{label5.2}
{\cal L}_{\rm Th}(x) &=&
\bar{\psi}(x)i\gamma^{\mu}\partial_{\mu}\psi(x) -
\frac{1}{2}\,g\,\bar{\psi}(x)\gamma^{\mu}\psi(x)\bar{\psi}(x)
\gamma_{\mu}\psi(x)\nonumber\\ &&+(Z_2 -
1)\,\bar{\psi}(x)i\gamma^{\mu}\partial_{\mu}\psi(x) -
\frac{1}{2}\,g\,(Z_1 -
1)\,\bar{\psi}(x)\gamma^{\mu}\psi(x)\bar{\psi}(x)\gamma_{\mu}\psi(x)
=\nonumber\\ &=&Z_2\,\bar{\psi}(x)i\gamma^{\mu}\partial_{\mu}\psi(x) -
\frac{1}{2}\,g\,Z_1\,\bar{\psi}(x) \gamma^{\mu}\psi(x)\bar{\psi}(x)
\gamma_{\mu}\psi(x),
\end{eqnarray}
where $Z_1$ and $Z_2$ are the renormalization constants of the
coupling constant and the wave function of the fermion field.

The renormalized fermionic field operator $\psi(x)$ and the coupling
constant $g$ are related to {\it bare} quantities by the relations
\cite{JC84}
\begin{eqnarray}\label{label5.3}
\psi_0(x) &=& Z^{1/2}_2\,\psi(x),\nonumber\\
g_0 &=& Z_1Z^{-2}_2\,g.
\end{eqnarray}
For the correlation functions of massless Thirring fermions the
renormalizability of the massless Thirring model means the possibility
to replace the infrared cut--off $\mu$ and the ultra--violet cut--off
$\Lambda$ by a finite scale $M$ by means of the renormalization
constants $Z_1$ and $Z_2$.

According to the general theory of renormalization \cite{JC84}, the
renormalization constants $Z_1$ and $Z_2$ depend on the renormalized
quantities $g$, the infrared scale $\mu$, the ultra--violet scale
$\Lambda$ and the finite scale $M$.  As has been shown above the Green
functions and left--right fermion density correlation functions do not
depend on the infrared cut--off.  Therefore, we can omit it. This
defines the renormalization constants as follows
\begin{eqnarray}\label{label5.4}
Z_1 &=& Z_1(g, M;\Lambda),\nonumber\\ Z_2 &=& Z_2(g, M;\Lambda).
\end{eqnarray}
For the analysis of the feasibility of the replacement $\Lambda \to M$
it is convenient to introduce the following notations
\begin{eqnarray}\label{label5.5}
\hspace{-0.3in}G^{(0)}(x_1,\ldots,x_n;y_1,\ldots,y_n) &=&
\Lambda^n\,G^{(0)}(d_{(\bar{\psi}\psi)}(g_0);\Lambda x_1,\ldots,\Lambda x_n;
\Lambda y_1,\ldots,\Lambda y_n),\nonumber\\
\hspace{-0.3in}C^{(0)}(x_1,\ldots,x_n;y_1,\ldots,y_n) &=&
\Lambda^{2n}\,C^{(0)}(d_{ (\bar{\psi}\psi)^2}(g_0);\Lambda x_1,\ldots,\Lambda
x_n;\Lambda y_1,\ldots,\Lambda y_n).
\end{eqnarray}
The transition to a finite scale $M$ changes the functions
(\ref{label5.5}) as follows
\begin{eqnarray}\label{label5.6}
&&G^{(0)}(x_1,\ldots,x_n;y_1,\ldots,y_n) 
=\nonumber\\
&&\hspace{0.5cm}=\Bigg(\frac{\Lambda}{M}\Bigg)^{\textstyle -
  2nd_{(\bar{\psi}\psi)}(g)}\,M^n\,G^{(0)}(d_{(\bar{\psi}\psi)}(g_0);M x_1,\ldots,M
x_n; M y_1,\ldots,M y_n),\nonumber\\
&&C^{(0)}(x_1,\ldots,x_n;y_1,\ldots,y_n) 
=\nonumber\\
&&\hspace{0.5cm}=\Bigg(\frac{\Lambda}{M}\Bigg)^{\textstyle -
4nd_{(\bar{\psi}\psi)^2}(g)}\,M^{2n}\,C^{(0)}(d_{(\bar{\psi}\psi)^2}(g_0); M
x_1,\ldots,M x_n;M y_1,\ldots,M y_n).
\end{eqnarray}
The renormalized correlation functions are related to the {\it bare}
ones by the relations \cite{JC84}:
\begin{eqnarray}\label{label5.7}
  &&G^{(r)}(x_1,\ldots,x_n;y_1,\ldots,y_n) 
  =Z^{-n}_2\,G^{(0)}(x_1,\ldots,x_n;y_1,\ldots,y_n)=\nonumber\\
  &&=Z^{-n}_2\Bigg(\frac{\Lambda}{M}\Bigg)^{\textstyle\!
    - 2nd_{(\bar{\psi}\psi)}(g)}\!M^n\,G^{(0)}(d_{(\bar{\psi}\psi)}(Z_1 Z^{-2}_2 g); M
  x_1,\ldots,M x_n; M y_1,\ldots,M y_n),\nonumber\\
  &&C^{(r)}(x_1,\ldots,x_n;y_1,\ldots,y_n) 
  =Z^{-2n}_2 C^{(0)}(x_1,\ldots,x_n;y_1,\ldots,y_n)= 
  \nonumber\\
  &&=Z^{-2n}_2\Bigg(\frac{\Lambda}{M}\Bigg)^{\textstyle\!
    - 4nd_{(\bar{\psi}\psi)^2}(g)}M^{2n}\,C^{(0)}(d_{(\bar{\psi}\psi)^2}(Z_1 Z^{-2}_2 g); M
  x_1,\ldots,M x_n;M y_1,\ldots,M y_n). \nonumber\\
\end{eqnarray}
Renormalizability demands the relations
\begin{eqnarray}\label{label5.8}
\hspace{-0.3in}G^{(r)}(x_1,\ldots,x_n;y_1,\ldots,y_n) &=&
M^n\,G^{(r)}(d_{(\bar{\psi}\psi)}(g); M x_1,\ldots,M x_n;M y_1,\ldots,M
y_n),\nonumber\\
\hspace{-0.3in}C^{(r)}(x_1,\ldots,x_n;y_1,\ldots,y_n) &=&
M^{2n}\,C^{(r)}(d_{(\bar{\psi}\psi)^2}(g); M x_1,\ldots,M x_n;M y_1,\ldots,M
y_n),
\end{eqnarray}
which impose constraints on the dynamical dimensions and
renormalization constants
\begin{eqnarray}\label{label5.9}
d_{(\bar{\psi}\psi)}(g) &=& d_{(\bar{\psi}\psi)}(Z_1Z^{-2}_2
g),\nonumber\\
d_{(\bar{\psi}\psi)^2}(g) &=&
d_{(\bar{\psi}\psi)^2}(Z_1Z^{-2}_2 g)
\end{eqnarray}
and 
\begin{eqnarray}\label{label5.10}
Z^{-1}_2\,\Bigg(\frac{\Lambda}{M}\Bigg)^{\textstyle - 2 d_{
(\bar{\psi}\psi)}(g)} &=&
Z^{-1}_2\,\Bigg(\frac{\Lambda}{M}\Bigg)^{\textstyle - 2 d_{
(\bar{\psi}\psi)^2}(g)} = 1.
\end{eqnarray}
The constraints (\ref{label5.9}) on the dynamical dimensions are
fulfilled only if the renormalization constants are related by
\begin{eqnarray}\label{label5.11}
Z_1 = Z^2_2.
\end{eqnarray}
The important consequence of this relation is that the coupling
constant $g$ of the massless Thirring model is unrenormalized, i.e.
\begin{eqnarray}\label{label5.12}
g_0 = g.
\end{eqnarray}
This also implies that the Gell--Mann--Low $\beta$--function, defined
by \cite{JC84}
\begin{eqnarray}\label{label5.13}
M\frac{dg}{dM} = \beta(g, M),
\end{eqnarray}
should vanish, since $g$ is equal to $g_0$, which does not depend on
$M$, i.e. $\beta(g, M) = 0$. Our observation concerning the
unrenormalizability of the coupling constant, $g_0 = g$, is supported
by the results obtained in \cite{MTM} for the massive Thirring
model.

The constraint (\ref{label5.10}) is fulfilled only for $d_{
  (\bar{\psi}\psi)}(g) = d_{ (\bar{\psi}\psi)^2}(g)$. In this case the
dependence of the $2n$--point causal Green functions and the
$2n$--point correlation functions of left--right fermion densities on
the ultra--violet cut--off $\Lambda$ can be simultaneously removed by
renormalization of the wave function of the massless Thirring fermion
fields.  This means the massless Thirring model is non--perturbative
renormalizable.

\section{Conclusion}
\setcounter{equation}{0}

We have found the most general expressions for the causal two--point
Green function and the two--point correlation function of left--right
fermion densities with dynamical dimensions parameterised by two
parameters.  The region of variation of these parameters is restricted
by the positive definiteness of the norms of the wave functions of the
states related to the components of the fermion vector current (see
Appendix B).

Our expressions incorporate those obtained by Hagen, Klaiber and
within the path--integral approach \cite{CH67}--\cite{RO81}. Indeed,
for Hagen's parameterisation of the functional determinant with the
parameters $\bar{\xi} = \xi$ and $\bar{\eta} = 1$ the dynamical
dimensions $d_{\bar{\psi}\psi}(g)$ and $d_{(\bar{\psi}\psi)^2}(g)$
take the form
\begin{eqnarray}\label{label6.1}
  d_{\bar{\psi}\psi}(g) = \frac{g^2}{2\pi^2}\,\frac{1}{\displaystyle \Big(1 +
    \xi\,\frac{g}{\pi}\Big)\Big(1 -
    \eta\,\frac{g}{\pi}\Big)} \quad,\quad  
  d_{(\bar{\psi}\psi)^2}(g) = - \frac{g}{2\pi}\, \frac{1}{\displaystyle 1 +
    \xi\,\frac{g}{\pi}}.
\end{eqnarray}
For $\xi = 1$ we get
\begin{eqnarray}\label{label6.2}
  d_{\bar{\psi}\psi}(g) = \frac{g^2}{2\pi^2}\,\frac{1}{\displaystyle 1 +
    \frac{g}{\pi}} \quad,\quad  
  d_{(\bar{\psi}\psi)^2}(g) = - \frac{g}{2\pi}\, \frac{1}{\displaystyle 1 +
    \frac{g}{\pi}}.
\end{eqnarray}
These are dynamical dimensions of the Green functions and correlation
functions of left--right fermion densities obtained by Klaiber
\cite{BK67} and within the path--integral approach
\cite{MF82}--\cite{RO81}.

We have shown that dynamical dimensions $d_{\bar{\psi}\psi}(g)$ and
$d_{(\bar{\psi}\psi)^2}(g)$ can be made equal. This fixes the
parameter $\bar{\eta}$ in terms of the parameter $\bar{\xi}$ and gives
the dynamical dimension of the massless Thirring fermion fields equal
to
\begin{eqnarray}\label{label6.3}
  d_{\bar{\psi}\psi}(g) = 
  d_{(\bar{\psi}\psi)^2}(g) = d_{\psi}(g) = - \frac{g}{2\pi}\, \frac{1}{\displaystyle 1 +
    \bar{\xi}\,\frac{g}{\pi}}.
\end{eqnarray}
As has been pointed out by Jackiw \cite{RJ71}, the inequality of
dynamical dimensions of fermion fields obtained from different
correlation functions is the problem of 1+1--dimensional quantum field
theories.  The equality of the dynamical dimensions
$d_{\bar{\psi}\psi}(g)$ and $d_{(\bar{\psi}\psi)^2}(g)$ is not
suppressed by the positive definiteness of the norms of the wave
functions of the states related to the components of the vector
currents. A positive definiteness of the norms of the wave functions
of these states imposes some constraints on the region of variation of
the parameters $\bar{\eta}$ and $\bar{\xi}$, demanding the parameter
$1 + \bar{\xi}\,g/\pi$ to be negative, i.e. $1 + \bar{\xi}\,g/\pi <
0$.

This makes the massless Thirring model renormalizable in the sense
that the dependence of correlation functions of Thirring fermion
fields on the ultra--violet cut--off can be removed by renormalization
of the wave function of Thirring fermion fields only. We have
corroborated this assertion within the standard renormalization
procedure.

From the constraint $ - g\,(1 + \bar{\xi}\,g/\pi) > 0$ there follows
that the coupling constant $\beta^2$ of the sine--Gordon model is of
order $\beta^2 \sim 8\pi$. A behaviour and renormalizability of the
sine--Gordon model for the coupling constants $\beta^2 \sim 8\pi$ has
been investigated in \cite{SGM}.

We are grateful to Manfried Faber for numerous helpful discussions.

\section*{Appendix A: On the parameterisation of the functional
  determinant ${\rm Det}(i\hat{\partial} + \hat{A})$}
\renewcommand{\theequation}{A-\arabic{equation}}
\setcounter{equation}{0}

The result of the calculation of the functional determinant ${\rm
  Det}(i\hat{\partial} + \hat{A})$ is related to the vacuum
expectation value $\langle j^{\mu}(x)\rangle$ of the vector current
$j^{\mu}(x) = \bar{\psi}(x)\gamma^{\mu}\psi(x)$. Using
(\ref{label2.4}), the vacuum expectation value of
the vector current can be defined by
\begin{eqnarray}\label{labelA.1}
  \langle j^{\mu}(x)\rangle &=& \frac{1}{i}\frac{\delta}{\delta
    A_{\mu}(x)}{\ell n}Z^{(0)}_{\rm th}[A,J,\bar{J}]\Big|_{\bar{J} = J = 0} = \nonumber\\
&=&\frac{1}{i}
\frac{\delta}{\delta A_{\mu}(x)}{\ell n} {\rm
    Det}(i\hat{\partial} + \hat{A})= \int d^2y\,D^{\mu\nu}(x - y)\,A_{\nu}(y),
\end{eqnarray}
where $D^{\mu\nu}(x - y)$ is given by (\ref{label2.7}) and
parameterised by two parameters $\bar{\xi}$ and
$\bar{\eta}$. Hence, the calculation of the vacuum expectation value
of the vector current should show how many parameters one can use for
the parameterisation of the Green function $D^{\mu\nu}(x - y)$ or the
functional determinant ${\rm Det}(i\hat{\partial} + \hat{A})$ as well.
According to Hagen \cite{CH67}, $\langle j^{\mu}(x)\rangle$ can be
determined by
\begin{eqnarray}\label{labelA.2}
\langle j^{\mu}(x)\rangle &=& \lim_{y \to x}{\rm tr}\Big\{i
G(x,y)_A\gamma^{\mu}\exp\,i\int^y_xdz^{\nu}(a\,A_{\nu}(z) +
b\,\gamma^5\,A_{5\nu}(z))\Big\}, 
\end{eqnarray}
where $a$ and $b$ are parameters and $A^{\nu}_5(z) =
-\,\varepsilon^{\nu\beta}A_{\beta}(z)$. The fermion Green function
$G(y,x)_A$ is given by (\ref{label2.9}). The requirement of covariance
relates the parameters $a$ and $b$. This provides the parameterisation
of the functional determinant ${\rm Det}(i\hat{\partial} + \hat{A})$
by one  parameter. In Hagen's notation this is the parameter
$\xi$.

In order to show that the functional determinant ${\rm
  Det}(i\hat{\partial} + \hat{A})$ can be parameterised by two
parameters we propose to define the vacuum expectation value
of the vector current (\ref{labelA.2}) as follows
\begin{eqnarray}\label{labelA.3}
 \hspace{-0.3in}  &&\langle j^{\mu}(x)\rangle = \lim_{y \to x}{\rm
  tr}\Big\{i G(x,y)_A\gamma^{\mu}\exp\,i\int^y_xdz^{\nu}\Big(a\,A_{\nu}(z) +
  b\,\gamma^5\,A_{5\nu}(z)\nonumber\\
  \hspace{-0.3in}  &&+ c\int d^2t\,\frac{\partial}{\partial
  t^{\nu}}\frac{\partial}{\partial t_{\beta}}\Delta(z -
  t;\mu)\,A_{\beta}(t) + d\,\gamma^5 \int d^2t\,\frac{\partial}{\partial
  t^{\nu}}\frac{\partial}{\partial t_{\beta}}\Delta(z -
  t;\mu)\,A_{5\beta}(t)\Big) \Big\}, 
\end{eqnarray}
where $c$ and $d$ are additional parameters and $\Delta(z -
t;\mu)$ is determined by (\ref{label2.8}). Under the gauge
transformation $A_{\nu} \to A\,'_{\nu} = A_{\nu} + \partial_{\nu}\phi$
the third term in (\ref{labelA.3}) behaves like the first one, whereas
the fourth one is gauge invariant.

The vacuum expectation value of the vector current can be transcribed
into the form
\begin{eqnarray}\label{labelA.4}
 && \langle j^\mu(x)\rangle = \lim_{y\rightarrow x}\frac{i}{2\pi}
\frac{(x - y)_\rho}
  {(x - y)^2 - i0}\,{\rm tr}\Big\{\gamma^\rho\exp\,(-i(g^{\alpha\beta} -
  \varepsilon^{\alpha\beta}\gamma^5)\nonumber\\
&&\times\int d^2z\,\frac{\partial}{\partial z^\alpha}[\Delta(x - z;\mu) -
 \Delta(y - z;\mu)]A_\beta(z))\,
  \gamma^\mu \exp\,i\int_x^y dz^\nu\,\Big(aA_\nu(z) +
  b\gamma^5A_{5\nu}(z)\nonumber\\
&&+ c\int d^2t\,\frac{\partial}{\partial t^\nu}\frac{\partial}{\partial
  t_\beta}\Delta(z - t;\mu)A_\beta(t) + d\gamma^5\int
d^2t\,\frac{\partial}{\partial t^\nu}\frac{\partial}{\partial
  t_\beta}\Delta(z - t;\mu) A_{5\beta}(t)\Big)\Big\}.
\end{eqnarray}
For the calculation of the r.h.s. of (\ref{labelA.4}) we apply the
spatial--point--slitting technique. We set $y^0=x^0$ and
$y^1=x^1\pm\epsilon$, taking the limit $\epsilon \to 0$. This gives
\begin{eqnarray}\label{labelA.5}
  \langle j^\mu(x)\rangle&=&\lim_{\epsilon\rightarrow0}\frac{i}{2\pi}
  \frac{1}{\mp\epsilon}
  \mbox{tr}\Big\{\gamma^1\Big[1\mp i\epsilon(g^{\alpha\beta} - 
  \varepsilon^{\alpha\beta}\gamma^5)
  \frac{\partial}{\partial x^1}\frac{\partial}{\partial x^\alpha}\int d^2z\,
  \Delta(x - z;\mu)A_\beta(z)\Big]\nonumber\\
  &&\times\gamma^\mu\Big[1\pm i\epsilon\Big(aA_1(x) + b\gamma^5 
  A_{51}(x) + c\int d^2t\,
  \frac{\partial}{\partial t^1}\frac{\partial}{\partial t_\beta}
  \Delta(x - t;\mu)A_\beta(t)\nonumber\\
  &&+ d\gamma^5\int d^2t\,\frac{\partial}{\partial t^1}
  \frac{\partial}{\partial t_\beta}
  \Delta(x - t;\mu)A_{5\beta}(t)
  \Big)\Big]\Big\} = \nonumber\\
  &=&\mp\lim_{\epsilon\rightarrow0}\frac{ig^{1\mu}}{\pi\epsilon} 
  +\lim_{\epsilon\rightarrow0}\frac{i}{2\pi\epsilon}
  \Big[ i\epsilon(2g^{1\mu}g^{\alpha\beta} + 2\varepsilon^{1\mu}
  \varepsilon^{\alpha\beta})
  \frac{\partial}{\partial x^1}\frac{\partial}{\partial x^\alpha}\int d^2z\,
  \Delta(x - z;\mu)A_\beta(z) \nonumber\\
  &&\mp i\epsilon\Bigl(2ag^{1\mu}A_1(x) + 2b\varepsilon^{1\mu}A_{51}(x) + 
  2cg^{1\mu}\int d^2t\,
  \frac{\partial}{\partial t^1}\frac{\partial}{\partial t_\beta}
  \Delta(x - t;\mu)A_\beta(t)\nonumber\\
  &&+2d\,\varepsilon^{1\mu}\int d^2t\,\frac{\partial}{\partial t^1}
  \frac{\partial}{\partial t_\beta}\Delta(x - t;\mu)
  A_{5\beta}(t)\Big)\Big],
\end{eqnarray}
Taking the symmetric limit we get
\begin{eqnarray}\label{labelA.6}
\langle j^\mu(x)\rangle&=&\frac{1}{\pi}\Big[-(g^{1\mu}g^{\alpha\beta} +
 \varepsilon^{1\mu}\varepsilon^{\alpha\beta})
  \frac{\partial}{\partial x^1}\frac{\partial}{\partial x^\alpha}\int d^2z\,
\Delta(x - z;\mu)A_\beta(z)\nonumber\\
&&+\Bigl(ag^{1\mu}A_1(x) + b\varepsilon^{1\mu}A_{51}(x) + cg^{1\mu}\int d^2t\,
  \frac{\partial}{\partial t^1}\frac{\partial}{\partial t_\beta}\Delta(x - t;\mu)
A_\beta(t)\nonumber\\
&&+ d\varepsilon^{1\mu}\int d^2t\,\frac{\partial}{\partial t^1}
\frac{\partial}{\partial t_\beta}
\Delta(x - t;\mu)
  A_{5\beta}(t)\Bigr)\Big].
\end{eqnarray}
The components of the current are equal to
\begin{eqnarray}\label{labelA.7}
  \langle j^0(x)\rangle&=&\frac{1}{\pi}\varepsilon^{\alpha\beta}
  \frac{\partial}{\partial x^1}\frac{\partial}
  {\partial x^\alpha}\int d^2z\,\Delta(x - z;\mu)A_\beta(z)\nonumber\\ 
  &&-\frac{b}{\pi}A_{51}(x) - \frac{d}{\pi}\int
  d^2t\,\frac{\partial}{\partial t^1}\frac{\partial} {\partial
  t_\beta}\Delta(x - t;\mu)A_{5\beta}(t),\nonumber\\
  \langle j^1(x)\rangle&=&\frac{1}{\pi}\frac{\partial}{\partial
  x^1}\frac{\partial}{\partial x_\alpha} \int d^2z\,\Delta(x -
  z;\mu)A_\alpha(z)\nonumber\\
  &&-\frac{a}{\pi}\,A_{1}(x) - \frac{c}{\pi}\int
  d^2t\,\frac{\partial}{\partial t^1}\frac{\partial} {\partial
  t_\beta}\Delta(x - t;\mu)A_\beta(t).
\end{eqnarray}
Using $A_{5\mu}=-\varepsilon_{\mu\nu}A^\nu$ and $\Box\Delta(x - y;\mu)
= \delta^{(2)}(x - y)$ the zero component can be transcribed into the
form
\begin{eqnarray}\label{labelA.8}
  \langle j^0(x)\rangle&=&\frac{1}{\pi}\frac{\partial}{\partial x^1}
  \frac{\partial}{\partial x^0}\int d^2z\,\Delta(x - z;\mu)A_1(z) 
  -\frac{1}{\pi}\frac{\partial}{\partial x^1}
  \frac{\partial}{\partial x^1}\int d^2z\,
  \Delta(x - z;\mu)A_0(z) \nonumber\\
  &&+\frac{b}{\pi}\,A^0(x) +  \frac{d}{\pi}\,\int d^2t
  \frac{\partial}{\partial t^1}\frac{\partial}
  {\partial t^0}\Delta(x - t;\mu)A_1(t) \nonumber\\
  &&- \frac{d}{\pi}\int d^2t\frac{\partial}{\partial t^1}\frac{\partial}
  {\partial t^1}\Delta(x - t;\mu)A_0(t) = \nonumber\\
  &=&-\frac{1}{\pi}\,\frac{\partial}{\partial x^0}\frac{\partial}{\partial x_\mu}
  \int d^2z\,\Delta(x - z;\mu)A_\mu(z) 
  + \frac{d}{\pi}\,A^0(x) \nonumber\\
  &&+ \frac{b + 1}{\pi}\,A^0(x) - \frac{d}{\pi}\int d^2t\frac{\partial}{\partial t^0}
  \frac{\partial}{\partial t_\mu}\Delta(x - t;\mu)A_\mu(t) = \nonumber\\ 
  &=&-\frac{1 + d}{\pi}\frac{\partial}{\partial x_0}\frac{\partial}{\partial x_\mu}
  \int d^2z\,\Delta(x - z;\mu)A_\mu(z) 
  + \frac{b + d + 1}{\pi}\,A^0(x).
\end{eqnarray}
Comparing the time component with the spatial one, given by
\begin{eqnarray}\label{labelA.9}
  \langle j^1(x)\rangle&=&\frac{c - 1}{\pi}\,\frac{\partial}{\partial x_1}
\frac{\partial}{\partial x_\mu}
  \int d^2z\,\Delta(x - z;\mu)A_\mu(z) + \frac{a}{\pi}\,A^{1}(x),
\end{eqnarray}
we obtain that the covariance of the vacuum expectation value of the
vector current takes place for $c = - d$ and $b + d + 1 = a$ only.

Thus, the vacuum expectation value of the vector current is
\begin{eqnarray}\label{labelA.10}
  \langle j^\mu(x)\rangle &=& \frac{\bar\xi}{\pi}\, A^{\mu}(x) - 
  \frac{\bar\eta}{\pi}\,
  \frac{\partial}{\partial x_\mu}\frac{\partial}{\partial x_\nu}
  \int d^2z\,\Delta(x - z;\mu)A_\nu(z) \nonumber\\
  &=& \int d^2y\,D^{\mu\nu}(x - y)\,
  A_{\nu}(y),
\end{eqnarray}
where $\bar\eta$ and $\bar\xi$ are parameters related to the
parameters $a$, $b$, $c$ and $d$ as $\bar{\xi} = a$ and $\bar{\eta} =
1 - c$.  The vacuum expectation value of the vector current, given
by (\ref{labelA.10}), supports the possibility to parameterise the
functional determinant ${\rm Det}(i\hat{\partial} + \hat{A)}$ as well
as the Green function $D^{\mu\nu}(x - y)$ by two parameters
(\ref{label2.7}).

\section*{Appendix B: Constraints on the parameters $\bar{\xi}$ and
  $\bar{\eta}$ from the norms of the wave functions of the states
  related to the components of the vector current}
\renewcommand{\theequation}{B-\arabic{equation}}
\setcounter{equation}{0}

The dependence of the functional determinant ${\rm
  Det}(i\hat{\partial} + \hat{A)}$ on two parameters leads to the
dependence of the two--point correlation function $\langle 0|{\rm
  T}(j^{\mu}(x)j^{\nu}(y))|0\rangle$ on these parameters.  Following
Johnson \cite{KJ61}, for the vacuum expectation value $\langle 0|{\rm
  T}(j^{\mu}(x)j^{\nu}(y))|0\rangle$ we get
\begin{eqnarray}\label{labelB.1}
  i\langle 0|{\rm
    T}(j^{\mu}(x)j^{\nu}(y))|0\rangle &=&  
  - \frac{\bar{\eta}}{\pi}\,\frac{1}{\displaystyle \Big(1 +
    \bar{\xi}\,\frac{g}{\pi}\Big)\Big(1 +
    (\bar{\xi} - \bar{\eta})\,\frac{g}{\pi}\Big)}\,\frac{\partial}{\partial
    x_{\mu}}\frac{\partial}{\partial x_{\nu}}\Delta(x - y;\mu)\nonumber\\
&&+ \frac{\bar{\eta}}{\pi}\,\frac{1}{\displaystyle \Big(1 +
    \bar{\xi}\,\frac{g}{\pi}\Big)\Big(1 +
    (\bar{\xi} - \bar{\eta})\,\frac{g}{\pi}\Big)}\,g^{\mu0}g^{\nu0}\,\delta^{(2)}(x - y).
\end{eqnarray}
This gives the following expressions for the vacuum expectation values
$\langle 0|j^0(x)j^0(y)|0\rangle$ and $\langle
0|j^1(x)j^1(y)|0\rangle$
\begin{eqnarray}\label{labelB.2}
  \langle 0|j^0(x)j^0(y)|0\rangle &=&  - \frac{\bar{\eta}}{\pi}\,\frac{1}{\displaystyle \Big(1 +
    \bar{\xi}\,\frac{g}{\pi}\Big)\Big(1 +
    (\bar{\xi} - \bar{\eta})\,\frac{g}{\pi}\Big)}\,\Big(\frac{\partial}{\partial x^1}\Big)^2 
  D^{(+)}(x - y)\nonumber\\
  \langle 0|j^1(x)j^1(y)|0\rangle &=& - \frac{\bar{\eta}}{\pi}\,\frac{1}{\displaystyle \Big(1 +
    \bar{\xi}\,\frac{g}{\pi}\Big)\Big(1 +
    (\bar{\xi} - \bar{\eta})\,\frac{g}{\pi}\Big)}\,
  \Big(\frac{\partial}{\partial x^1}\Big)^2 D^{(+)}(x - y),
\end{eqnarray}
where $D^{(\pm)}(x - y)$ are the Wightman functions given by
\begin{eqnarray}\label{labelB.3}
  D^{(\pm)}(x - y) = \int \frac{d^2k}{(2\pi)^2}\,2\pi\theta(k^0)\delta(k^2)\,
e^{\textstyle \,\mp ik\cdot (x - y)}
\end{eqnarray}
We have taken into account that
\begin{eqnarray}\label{labelB.4}
  \Delta(x - y;\mu) = i\theta(x^0 -y^0)\, D^{(+)}(x - y) + i\theta(y^0 -x^0)\, D^{(-)}(x - y).
\end{eqnarray}
According to Wightman and Streater \cite{AW80} and Coleman
\cite{SC73}, we can define the wave functions of the states
\begin{eqnarray}\label{labelB.5}
 |h;j^0\rangle &=& \int d^2x\,h(x)\,j^0(x)|0\rangle,\nonumber\\
|h;j^1\rangle &=& \int d^2x\,h(x)\,j^1(x)|0\rangle,
\end{eqnarray}
where $h(x)$ is the test function from the Schwartz class $h(x)\in
{\cal S}({\mathbb{R}^{\,2}})$ \cite{AW80}.

The norms of the states (\ref{labelB.5}) are equal to \cite{AW80,SC73}
\begin{eqnarray}\label{labelB.6}
  &&\langle j^0;h|h;j^0\rangle = \int\!\!\!\int d^2x d^2y\,h^*(x)\,
  \langle 0| j^0(x)j^0(y)|0\rangle\, h(y) 
  =\nonumber\\
  &&= \frac{\bar{\eta}}{\pi}\,\frac{1}{\displaystyle \Big(1 +
    \bar{\xi}\,\frac{g}{\pi}\Big)\Big(1 +
    (\bar{\xi} - \bar{\eta})\,\frac{g}{\pi}\Big)}\int 
  \frac{d^2k}{(2\pi)^2}\,2\pi\,
  (k^0)^2\theta(k^0)\delta(k^2)\,|\tilde{h}(k)|^2,\nonumber\\
  &&\langle j^1;h|h;j^1\rangle = \frac{\bar{\eta}}{\pi}\,
  \frac{1}{\displaystyle \Big(1 +
    \bar{\xi}\,\frac{g}{\pi}\Big)\Big(1 +
    (\bar{\xi} - \bar{\eta})\,\frac{g}{\pi}\Big)}\int\!\!\!\int 
  d^2x d^2y\,h^*(x)\,
  \langle 0 | j^1(x)j^1(y)|0\rangle\, h(y)
  = \nonumber\\
  &&=  \frac{\bar{\eta}}{\pi}\,\frac{1}{\displaystyle \Big(1 +
    \bar{\xi}\,\frac{g}{\pi}\Big)\Big(1 +
    (\bar{\xi} - \bar{\eta})\,\frac{g}{\pi}\Big)}\int \frac{d^2k}{(2\pi)^2}
  \,2\pi\,
  (k^0)^2\theta(k^0)\delta(k^2)\,|\tilde{h}(k)|^2,
\end{eqnarray}
where $\tilde{h}(k)$ is the Fourier transform of the test function
$h(x)$. Since the norms of the states (\ref{labelB.5}) should be
positive, we get the constraint
\begin{eqnarray}\label{labelB.7}
  \bar{\eta}\,\Big(1 +
    \bar{\xi}\,\frac{g}{\pi}\Big)\Big(1 +
    (\bar{\xi} - \bar{\eta})\,\frac{g}{\pi}\Big) > 0.
\end{eqnarray}
This assumes that $\bar{\eta} \neq 0$.  For $\bar{\eta}$, constrained
by the requirement of the renormalizability of the massless Thirring
model (\ref{label4.5}), the inequality (\ref{labelB.7}) reduces to the
form
\begin{eqnarray}\label{labelB.8}
  -\,g\,\Big(1 +
    \bar{\xi}\,\frac{g}{\pi}\Big) > 0.
\end{eqnarray}
This inequality is fulfilled for
\begin{eqnarray}\label{labelB.9}
  g > 0 \;,\; 1 + \bar{\xi}\,\frac{g}{\pi} < 0 \quad,\quad 
  g < 0 \;,\; 1 + \bar{\xi}\,\frac{g}{\pi} > 0.
\end{eqnarray}
Using the vacuum expectation value of the two--point correlation
function of the vector currents (\ref{labelB.1}) we can calculate the
equal--time commutator $[j^0(x),j^1(y)]_{x^0 = y^0}$ and the Schwinger
term. We get
\begin{eqnarray}\label{labelB.10}
  [j^0(x),j^1(y)]_{x^0 = y^0} &=&  -\,c\,i\frac{\partial}{\partial x^1}\delta(x^1 - y^1) =\nonumber\\
&=&  -\,\frac{\bar{\eta}}{\pi}\,\frac{1}{\displaystyle \Big(1 +
    \bar{\xi}\,\frac{g}{\pi}\Big)\Big(1 +
    (\bar{\xi} - \bar{\eta})\,\frac{g}{\pi}\Big)}\,i\frac{\partial}{\partial x^1}\delta(x^1 - y^1)
\end{eqnarray}
with the Schwinger term $c$ equal to
\begin{eqnarray}\label{labelB.11}
  c = \frac{\bar{\eta}}{\pi}\,\frac{1}{\displaystyle \Big(1 +
    \bar{\xi}\,\frac{g}{\pi}\Big)\Big(1 +
    (\bar{\xi} - \bar{\eta})\,\frac{g}{\pi}\Big)}.
\end{eqnarray}
Due to the constraint (\ref{labelB.7}) it is always positive. For
$\bar{\eta} = 1$ our expression (\ref{labelB.10}) for the equal--time
commutator coincides with that obtained by Hagen \cite{CH67}.

Using (\ref{labelB.10}) we can analyse the Bjorken--Johnson--Low (BJL)
limit for the Fourier transform of the two--point correlation function
of the vector currents \cite{BJL,HP73}. Following \cite{HP73}, we
consider the Fourier transform
\begin{eqnarray}\label{labelB.12}
  T_{\mu\nu}(q) = i\int d^2x\,e^{\textstyle\,iq\cdot x}\,\langle A|{\rm T}(j_{\mu}(x)j_{\nu}(0))|B\rangle,
\end{eqnarray}
where $q = (q^0,q^1)$ and $|A\rangle$ and $|B\rangle$ are quantum
states \cite{HP73}. In our case these are vacuum states $|A\rangle =
|B\rangle = |0\rangle$. This gives
\begin{eqnarray}\label{labelB.13}
  T_{\mu\nu}(q) = i\int d^2x\,e^{\textstyle\,iq\cdot x}\,\langle 0|{\rm T}(j_{\mu}(x)j_{\nu}(0))|0\rangle.
\end{eqnarray}
According to the BJL theorem \cite{BJL,HP73}, in the limit $q^0 \to
\infty$ the r.h.s of Eq.(\ref{labelB.12}) behaves as follows \cite{HP73}
\begin{eqnarray}\label{labelB.14}
  T_{\mu\nu}(q) &=& -\,\frac{1}{q^0}\int^{+\infty}_{-\infty}dx^1\,e^{\textstyle\,-iq^1x^1}
  \langle 0|[j_{\mu}(0,x^1),j_{\nu}(0)]|0\rangle\nonumber\\
&&- \frac{i}{(q^0)^2}\int^{+\infty}_{-\infty}dx^1\,
  e^{\textstyle\,-iq^1x^1}\langle 0|[\partial_0 j_{\mu}(0,x^1),j_{\nu}(0)]|0\rangle  + 
O\Big(\frac{1}{(q^0)^3}\Big)
\end{eqnarray}
For the time--space component of the two--point correlation function we get
\begin{eqnarray}\label{labelB.15}
 \hspace{-0.3in} T_{01}(q^0, q^1) &=& -\,\frac{1}{q^0}\int^{+\infty}_{-\infty}dx^1\,e^{\textstyle\,-iq^1x^1}
  \langle 0|[j_0(0,x^1),j_1(0)]|0\rangle\nonumber\\
\hspace{-0.3in}&&- \frac{i}{(q^0)^2}\int^{+\infty}_{-\infty}dx^1\,
  e^{\textstyle\,-iq^1x^1}\langle 0|[\partial_0 j_0(0,x^1),j_1(0)]|0\rangle  + 
O\Big(\frac{1}{(q^0)^3}\Big)
\end{eqnarray}
Using (\ref{labelB.10}) for the BJL limit of $T_{01}(q^0, q^1)$ we
obtain
\begin{eqnarray}\label{labelB.16}
  T_{01}(q^0, q^1) &=& \frac{q^1}{q^0}\,\frac{\bar{\eta}}{\pi}\,
\frac{1}{\displaystyle \Big(1 +
    \bar{\xi}\,\frac{g}{\pi}\Big)\Big(1 +
    (\bar{\xi} - \bar{\eta})\,\frac{g}{\pi}\Big)} + 
  O\Big(\frac{1}{(q^0)^3}\Big).
\end{eqnarray}
For $\bar{\eta} = 1$ this reproduces the result which can be obtained
using Hagen's solution \cite{CH67}.  One can show that due to
conservation of vector current the term proportional to $1/q^2_0$
vanishes.  The asymptotic behaviour of the Fourier transform of the
two--point correlation function of the vector currents places no
additional constraints on the parameters $\bar{\xi}$ and $\bar{\eta}$.

The inequality (\ref{labelB.7}) leads to the following interesting
consequences.  According to Coleman \cite{SC75}, the coupling constant
$\beta^2$ of the sine--Gordon model is related to the coupling
constant $g$ of the Thirring model as
\begin{eqnarray}\label{labelB.17}
  \frac{\beta^2}{8\pi}= \frac{1}{2} + d_{(\bar{\psi}\psi)^2}(g) = \frac{1}{2}\Big(1 - \frac{g}{\pi}\,
  \frac{1}{\displaystyle 1 + \bar{\xi}\,\frac{g}{\pi}}\Big).
\end{eqnarray}
Hence, for the constraint (\ref{labelB.9}) the coupling constant
$\beta^2$ is of order $\beta^2 \sim 8\pi$. A behaviour and
renormalizability of the sine--Gordon model for the coupling constants
$\beta^2 \sim 8\pi$ has been investigated in \cite{SGM}.


\begin{thebibliography}{9}
\bibitem{WT58} 
W. Thirring, 
Ann. Phys. (N.Y.) {\bf 3}, 91 (1958).
\bibitem{KJ61}
K. Johnson,
Nuovo Cim. {\bf 20}, 773 (1961).
\bibitem{FS62}
F. L. Scarf and J. Wess,
Nuovo Cim. {\bf 26}, 150 (1962).
\bibitem{CH67}
C. R. Hagen,
Nuovo Cim. B {\bf 51}, 169 (1967).
\bibitem{BK67}
B. Klaiber, 
in {\it Lectures in theoretical physics},
Lectures delivered at the Summer Institute for Theoretical Physics,
University of Colorado, Boulder, 1967, edited by A. Barut and
W. Brittin, Gordon and Breach, New York, 1968, Vol. X, 
part A, pp.141--176.
\bibitem{MF82}
K. Furuya, Re. E. Gamboa Saravi and F. A. Schaposnik,
Nucl. Phys. B {\bf 208}, 159 (1982).
\bibitem{NA85}
C. M. Na${\acute{\rm o}}$n,
Phys. Rev. D {\bf 31}, 2035 (1985).
\bibitem{RO81}
R. Roskies and F. A. Schaposnik,
Phys. Rev. D {\bf 23}, 558 (1981);
R. E. Gamboa Saravi, F. A. Schaposnik, and J. E. Solomin,
Nucl. Phys. B {\bf 153}, 112 (1979);
R. E. Gamboa Savari, M. A. Muschetti, F. A. Schaposnik, 
and J. E. Solomin,
Ann. of Phys. (N.Y.) {\bf 157}, 360 (1984);
O. Alvarez,
Nucl. Phys. B {\bf 238}, 61 (1984);
H. Dorn,
Phys. Lett. B {\bf 167}, 86 (1986).
\bibitem{KH86}
K. Harada, H. Kubota, and I. Tsutsui,
Phys. Lett. B {\bf 173}, 77 (1986).
\bibitem{RJ84}
R. Jackiw and R. Rajaraman,
Phys. Rev. Lett. {\bf 54}, 1219 (1985).
\bibitem{GC83}
G. A. Christos,
Z. Phys. C {\bf 18}, 155 (1983); 
Erratum Z. Phys. C {\bf 20}, 186 (1983);
A. Smailagic and R. E. Gamboa--Saravi,
Phys. Lett. B {\bf 192}, 145 (1987);
R. Banerjee,
Z. Phys. C {\bf 25}, 251 (1984);
T. Ikehashi,
Phys. Lett. B {\bf 313}, 103 (1993).
\bibitem{RJ71}
R. Jackiw,
Phys. Rev. D {\bf 3}, 2005 (1971).
\bibitem{JC84}
J. C. Collins,
in {\it Renormalization},
Cambridge University Press, Cambridge, 1984.
\bibitem{SC75}
S. Coleman,
Phys. Rev. D {\bf 11}, 2088 (1975).
\bibitem{SGM}
D. Amit, Y. Y. Goldschmidt, and G. Grinstein,
J. Phys. A {\bf 13}, 585 (1980);
K. Huang and J. Polonyi,
Int. Mod. Phys. A {\bf 6}, 409 (1991);
Zs. Gul$\acute{\rm a}$csi and  M. Gul$\acute{\rm a}$csi,
Adv. Phys. {\bf 47}, 1 (1998);
I. N$\acute{\rm a}$ndori, J. Polonyi, and K. Sailer,
Phys. Rev. D {\bf 63}, 045022 (2001); Philos. Mag. B {\bf 81},
1615 (2001); 
G. von Gersdorf and C. Wetterich,
Phys. Rev. B {\bf 64}, 054513 (2001);
I. N$\acute{\rm a}$ndori, K. Sailer, U. D. Jentschura, and
G. Soff,
J. Phys. G {\bf 28}, 607 (2002);
M. Faber and A. N. Ivanov,
J. Phys. A {\bf 36}, 7839 (2003) and references therein.
I. N$\acute{\rm a}$ndori, K. Sailer, U. D. Jentschura, and
G. Soff,
Phys. Rev. D {\bf 69}, 025004 (2004);
H. Bozkaya, M. Faber, A. N. Ivanov, M. Pitschmann, 
J. Phys. A: Math. Gen. {\bf 39}, 2177 (2006).
\bibitem{JS62}
J. Schwinger,
Phys. Rev. {\bf 128}, 2425 (1962).
\bibitem{MTM}
A. H. Mueller and T. L. Trueman,
Phys. Rev. D {\bf 4}, 1635 (1971);
M. Gomes and J. H. Lowenstein,
Nucl. Phys. B {\bf 45}, 252 (1972).
(2001).
\bibitem{AW80} 
R. F. Streater and A. S. Wightman,
in {\it PCT, spin and statistics},
Princeton University Press, Princeton and Oxford, Third Edition, 
1980.
\bibitem{SC73}
S. Coleman,
Comm. Math. Phys. {\bf 31}, 259 (1973). 
\bibitem{BJL}
J. D. Bjorken,
Phys. Rev. {\bf 148}, 1467 <81966);
K. Johnson and F. E. Low,
Progr. Theor. Phys., Suppl. {\bf 37--38}, 74 (1966).
\bibitem{HP73}
V. De Alfaro, S. Fubini, G. Furlan, and C. Rossetti,
in {\it Currents in hadronic physics}, North--Holland Publishing Co., 
Amsterdam $\,\bullet\,$ London, 1973, 
\end{thebibliography}
\end{document}